\magnification=1200
\tolerance=5000
\null
\hyphenpenalty=2000

\def\SH{{\cal H}}
\nopagenumbers

\vskip0.5truein
\centerline{\bf  A Search for 
Line Shape and Depth}
\centerline {\bf Variations in 51 Pegasi and $\tau$ Bo\"otis}
\vskip0.7truein

\centerline{Timothy M. Brown, Rubina Kotak}
\centerline{High Altitude Observatory/National Center for Atmospheric Research
\footnote{$^*$}{The National Center for Atmospheric Research is Sponsored by
the National Science Foundation}}
\centerline{3450 Mitchell Ln, P.O Box 3000, Boulder, CO 80307}
\centerline{brown@hao.ucar.edu, rubina@astro.lu.se}
\vskip10pt

\centerline{Scott D. Horner}
\centerline{Department of Astronomy \& Astrophysics, 
Pennsylvania State University}
\centerline{University Park, PA 16802}
\centerline{horner@astro.psu.edu}
\centerline{and}
\vskip10pt

\centerline{Edward J. Kennelly, Sylvain Korzennik, P. Nisenson 
\& Robert W. Noyes}
\centerline{Harvard-Smithsonian Center for Astrophysics, Cambridge MA 02138}
\centerline{tkennelly@cfa0.harvard.edu, sylvain@cfa0.harvard.edu,
nisenson@cfa0.harvard.edu}
\centerline{noyes@cfa0.harvard.edu}

\vskip0.7truein
Received \_\_\_\_\_\_\_\_\_\_\_\_\_\_\_\_\_\_\_\_\_\_\_\_\_\_\_\_\_\_\_\_\_
\_\_\_\_\_\_\_\_\_\_\_\_\_\_\_\_\_\_\_\_\_\_\_\_\_\_\_\_
\vfill\eject
\pageno=1
\footline={\hss\tenrm\folio\hss}

\centerline{\bf Abstract}

Spectroscopic observations of 51 Pegasi and $\tau$ Bo\"otis
show no periodic changes in the shapes of their line
profiles;
these results for 51 Peg are in significant conflict with those reported
by Gray and Hatzes (1997).
Our detection limits are small enough to rule out nonradial
pulsations as the cause of the variability in $\tau$ Boo,
but not in 51 Peg.
The absence of line shape changes is consistent with 
these stars' radial velocity variability arising from 
planetary-mass companions.

\vskip8pt
\parindent=0pt
Subject headings: line: profiles --- stars: individual (51 Peg, $\tau$ Boo) ---
stars: oscillations --- stars: planetary systems

\vskip12pt
\parindent=0pt
{\bf I. Introduction}
\parindent=20pt

Short-period
radial velocity variability of several tens of m s$^{-1}$
has recently been detected in several Sun-like stars,
suggesting the presence of planets with roughly jovian mass
in orbits smaller than 0.25 AU (Mayor \& Queloz 1995,
Butler {\it et al} 1997, Noyes {\it et al.} 1997).
Gray (1997) (henceforth G97) and Gray \& Hatzes (1997) 
(henceforth GH) challenged this planetary
interpretation, based on evidence for changes,
synchronous with the radial velocity variation,
in the spectrum of 51 Pegasi.
The parameters monitored by G97 and GH were
the shape of the $\lambda$6253 Fe I line (parameterized by its bisector
shape), and the ratio of the depths of
$\lambda$6252 V I and $\lambda$6253 Fe I.
Line profile changes would not be expected from
the gravitational attraction of
an orbiting body, but would be a natural consequence of
dynamic processes (perhaps nonradial pulsations) in the stellar
atmosphere.
It is important to know whether Gray's challenge is correct;
if it is, great efforts are being expended 
to explain planets that may not exist.
We report here the results of an independent spectroscopic study
of the stars 51 Peg and $\tau$ Boo,
searching for confirming evidence for periodic variations in
line shape or depth.
In order to get the most information out of our moderate-resolution
data we have introduced some new analysis techniques.
These and the related modeling of line profiles that have been
distorted by nonradial pulsations are described in Brown {\it et al.}
(1998);
here we shall report only our results.

\vskip12pt
\parindent=0pt
{\bf II. Observations and Modeling}
\parindent=20pt

We obtained the observations described here using the Advanced Fiber
Optic Echelle (AFOE) spectrograph at the 1.5 m Tillinghast telescope
at Mt. Hopkins, AZ.
This spectrograph is designed for precise radial velocity measurements,
and employs several techniques to assure a stable spectrograph point
spread function (Brown {\it et al.} 1994).
We observed 51 Peg and $\tau$ Boo with the AFOE configured to give
$\lambda/\delta\lambda = R \simeq 50000$.
A typical observation of each star consisted of a sequence of 3
sequential integrations, each lasting 10 m, and each giving S/N
of about 150 in the continuum near 600 nm.
Because they were intended for radial velocity studies, a large
majority of our observations were taken using an $I_2$ absorption
cell.
In these data, the stellar spectrum between 500 nm and 610 nm
is heavily blended with the $I_2$ spectrum,
therefore we did not use this wavelength range in any of our
line shape analysis.
In most cases we took only one 3-integration sequence each night,
but for a few nights we took 2 such sequences, one at the beginning
of the night and one at the end.
The observations of 51 Peg consist of 59 spectra, encompassing
20 sequences taken on 18 separate nights between 1995 Nov. 2 and
1996 Jul. 26.
Those of $\tau$ Boo consist of 90 spectra, 30 sequences, and
23 nights between 1996 Jun. 25 and 1997 Mar. 26.
The 56 m s$^{-1}$ radial velocity variation of 51 Peg 
(period 4.231 d) is
clearly evident in these data, as is the much larger
(468 m s$^{-1}$, 3.313 d) variation of $\tau$ Boo.

The AFOE's spectral resolution is not high enough to allow
useful measurements of spectrum line bisectors.
We therefore adopted a different method for characterizing the
shapes of spectrum lines -- a decomposition in terms of Hermite functions
$\SH_i(\lambda / \sigma)$, defined by
$$
\SH_n({\lambda \over \sigma}) \ = \ {N_n \over \sqrt \sigma}
\exp (-{\lambda^2 \over 2 \sigma^2} ) H_n({\lambda \over \sigma}) \ ,
\eqno(1)
$$
where $H_n(\lambda/\sigma)$ is a Hermite polynomial as defined by
Abromowitz and Stegun (1972), and $N_n$ is a normalization factor
given by $N_n \ = \ (2^n n! \sqrt \pi )^{-1/2}$.
The decomposition involved representing 
pieces of spectrum $I(\lambda)$ typically 1.5 nm wide
and centered on a fiducial wavelength $\lambda _f$
as follows:
$$
I(\lambda) \ = \ C[1 \ + \ S(\lambda - \lambda_f)] 
\left [ 1 \ - \ \sum \limits _j
D_j \left ( \SH_0( {\lambda - \lambda_{cj} \over \sigma} ) \ + \ 
\sum \limits _{i=3} ^5 h_i \SH_i ( {\lambda - \lambda_{cj} \over \sigma } )
\right ) \right ] \ ,
\eqno (2)
$$
where $C$ and $S$ describe the brightness and slope of the local continuum,
$D_j$ is the depth relative to continuum brightness of the $j^{\rm th}$
spectrum line, $\lambda_{cj}$ is the center wavelength of the $j^{\rm th}$
line, and $\sigma$ and the coefficients $h_i$ describe the width
and the departure from a pure Gaussian shape, assumed common
to all the lines. 
The parts of the line profile described by $h_1$ and $h_2$ have
been absorbed into $\lambda_c$ and $\sigma$, respectively.
Thus, a time series of the $h_i$ coefficients contains all of the
line shape information in our observations,
while the ratio of the $D$ parameters for the $\lambda$6252 V I and
the $\lambda$6253 Fe I lines reproduces G97's measurement
of line depth ratio.
In what follows,
we express the line shape coefficients $h_i$ as a percentage of $D_j$.
For the analysis described below, we fit 3 (for 51 Peg) or 4 (for $\tau$
Boo) sections of spectrum, each about 1.5 nm wide.
These wavelength ranges contained about 60 distinct spectrum lines 
for each star.
This use of multiple lines to gain noise immunity is justified when
seeking evidence for pulsations, since both theory and observation
(of, {\it e.g.}, $\delta$ Scuti stars -- see Kennelly {\it et al.} 1997)
indicate that pulsations modify line shapes in the same way for all but
very strong spectral lines.

To compare our results with those of G97 and GH,
we were obliged to model the line shapes emerging from pulsating
and rotating stars, computing both Hermite expansion coefficients $h_i$
and line bisector shape indices for a wide range of nonradial
pulsation parameters.
The modeling process was similar to that described by,
for example, Vogt \& Penrod (1983), by Hatzes (1996),
and by Schrijvers {\it et al.} (1997).
At each point on a fine grid sampling the observable stellar
hemisphere, we computed the line-of-sight component of velocity
resulting from the sum of rotation and pulsation.
The line profiles emerging from each point were then added together
with appropriate Doppler shifts and limb darkening to make up
the observed profile from the complete star.
We then convolved the disk-integrated profiles with instrumental
point spread functions to give those that would have been observed
with $R = 100000$ (to simulate G97's data) and with
$R=50000$ (to simulate the AFOE).
Last, we analyzed a series of such line profiles computed at various
phases of the oscillation cycle as if they were real data,
to determine the relationship between measures of bisector shape
and our own $h_i$ coefficients.

The principal conclusion from the modeling applies to slowly-rotating
stars such as 51 Peg and to fluid motions (such as those caused by
pulsation) that are sensibly constant over the range of depths
where the spectrum line is formed.
In this case,
our $h_3$ coefficient, the bisector span $S_b$, and bisector curvature 
$C_b$ all
measure essentially the same quantity:
the lowest-order antisymmetric distortion of the line profile that
is not a simple line shift.
For the residual flux values used by GH in their line bisector studies
(.85, .71, .48) and a central line depth of 0.6 (appropriate
for $\lambda$6253 Fe I),
the proportionality between $h_3$, $S_b$, and $C_b$ turns out to be
$$
\sigma \ h_3 \ \simeq 0.35 S_b \ \simeq \ -2.2 C_b \ ,
\eqno (3)
$$
where $\sigma$ is the Gaussian line width parameter in velocity units.
For processes ({\it e.g.} granulation) involving depth-dependent correlations
between velocity and temperature, this equivalence
need not hold.
But for the circumstances that most concern us,
$h_3$ is a good proxy for the bisector-based
measures of line shape oscillation.
We also computed the ratio between $h_3$ and the RV amplitude.
This ratio is quite variable, depending upon the mode spherical harmonic
degree $\ell$, the azimuthal order, the ratio of horizontal
to vertical velocities, and the inclination of the pulsation axis.
It is nevertheless possible to characterize the typical behavior for
each $\ell$,
which allowed us to estimate how large an $h_3$ variation to
expect, given a measured radial velocity variation.
Finally, we found that slowly rotating stars display
little change in line shape for a given radial velocity signal,
but that the shape changes grow more pronounced as $v \sin i$
starts to exceed the intrinsic line width.
Some of these conclusions disagree in important ways with 
the results of similar modeling by GH;
for more details, see Brown {\it et al.} (1998).

\vskip12pt
\parindent=0pt
{\bf III. Results of Time-Series Analysis}
\vskip8pt
{\it A. 51 Pegasi}
\parindent=20pt

We searched for periodic signals in the 51 Peg time series of $h_i$ 
and line depth ratio in two ways:
we computed periodograms of the time series, and we fit 
the amplitudes and phases of sinusoidal
functions with periods of 4.231 d (the radial velocity period)
and 2.575 d (the period of the largest bisector curvature signal
seen by GH).
No obvious narrow-band signals appear in the periodograms
at any frequency, for any of the $h_i$ or for the line strength
ratio.
We therefore used the periodograms
merely to provide estimates of the noise in each of the
parameters, 
using for this purpose the values within a frequency range of
$\pm$1 cycle per month about the periods of interest.
Table 1 shows the results of the sinusoid fitting, with error
estimates taken from the periodograms as just described.
For 51 Peg, only the $h_5$ amplitudes and the line ratio
amplitude for the 2.575 d period exceed zero by as much as
the estimated error $\sigma$;
the largest difference from zero is 1.7$\sigma$.
For the 4.231 d radial velocity period, we measure the variation
in line depth ratio to be a little less than half that reported
by G97, but the uncertainty is such that our result is
about equally consistent with G97's value or with no variation
at all.

What does the non-detection of an $h_3$ signal mean in physical terms?
Two kinds of answer are possible, as illustrated in Figure 1.
The line bisector curvature and the $h_3$ fitting coefficient
measure similar characteristics of spectral line asymmetry.
The ratio of these two quantities is therefore fairly well defined,
and is largely independent of model-based assumptions.
Thus one can estimate the $h_3$ amplitude that corresponds to the
45 m s$^{-1}$ curvature amplitude reported by GH.
The resulting amplitude is 2.54\%, which is inconsistent with the
measured $h_3$ amplitude with 10$\sigma$ significance.
By this measure, our observations are therefore strongly inconsistent
with those by GH.
On the other hand, if one assumes a nonradial pulsation with $\ell = 5$
and an observed radial velocity amplitude of 56 m s$^{-1}$,
then our modeling shows that under typical circumstances the
expected amplitudes of all $h_i$ coefficients should be below 0.5\%.
Such amplitudes are only marginally detectable in our data;
therefore, nonradial pulsations cannot be excluded by this test as the
cause of the radial velocity signal in 51 Peg.
The surface flow velocities associated with this pulsation would
peak at about 5 km s$^{-1}$, so the linearity assumption involved in
scaling our model results would probably be violated.
The dissipation associated with such near-sonic flows
would likely cause larger line profile variations than those we calculated,
however, so the actual limit on pulsation amplitude is probably smaller
than that given above.

\vskip12pt
\parindent=0pt
{\it B. $\tau$ Bo\"otis}
\parindent=20pt

The star $\tau$ Boo rotates more rapidly ($v \sin i = 15$ km s$^{-1}$),
than does 51 Peg 
and the amplitude of its sinusoidal radial velocity variation
(468 m s$^{-1}$) is a factor of 9 larger (Marcy {\it et al.} 1997).
Both of these circumstances make pulsations easier to detect in $\tau$ Boo
than in 51 Peg.
If the radial velocity variability of $\tau$ Boo were to arise from
nonradial pulsations, then our models show that the anticipated signals
in $h_3$, $h_4$, and $h_5$ should all have magnitudes of several percent.

We searched for such periodic signals in the line shapes of $\tau$ Boo
just as we did in those of 51 Peg, by examining the periodograms
and by fitting sinusoids to the known radial velocity period of 3.313 d.
The fitted values and their errors may be found in the last column of
Table 1.
Again, none of the fitted sinusoid amplitudes are significantly larger
than the estimated errors, so no line shape changes at the radial
velocity period have been detected.
Since $\tau$ Boo is hotter than 51 Peg, its $\lambda$6252 V I line is too
weak to allow a meaningful depth measurement.
For this reason, Table 1 gives no line depth ratio estimate for this star.

If nonradial pulsations were responsible for $\tau$ Boo's radial
velocity variation, our models lead us to expect an $h_3$ amplitude of
about 15\%.
Figure 2 shows the observed $h_3$ values for $\tau$ Boo phased to
the radial velocity period,
as well as a comparison between the fitted sinusoid and that expected from
the pulsation models.
In this case the results are unambiguous:
pulsations with large enough amplitude to cause the radial velocity signal
would also cause observable distortions of the line shapes, which
are not seen.

The $h_3$ periodogram for $\tau$ Boo shows a wide region of excess power
between 0.2 c d$^{-1}$ and 0.4 c d$^{-1}$. 
Its amplitude is small, however, (the rms signal associated with this
entire band is about 0.6\%) and the bandwidth is comparable to the 
estimated rotation
frequency of the star,
which is thought to be identical to the radial velocity frequency.
Perhaps this $h_3$ variation results from magnetic
active regions rotating across the disk;
it may also be associated with the excess radial velocity noise
noted in $\tau$ Boo by Butler {\it et al.} (1997).
In any case, it is too small in amplitude and covers too wide a
frequency band to be identified directly with the radial velocity signal.

\vskip12pt
\parindent=0pt
{\bf IV. Discussion}
\parindent=20pt

To summarize our conclusions: (a) We find no evidence for pulsations
in 51 Peg.
Our upper limits on line shape variations
are in significant conflict with the bisector curvature
measurements by GH.
(b) For the observed radial velocity signal, our models imply line shape
variations that are too small to be readily detected, either
by our methods or in GH's bisector data.
Thus, pulsations are not excluded as an explanation for
51 Peg's radial velocity variability.
(c) The radial velocity variability of $\tau$ Boo,
whose period is similar to that of 51 Peg,
does not arise from pulsations.

These conclusions have implications for the study of both extra-solar planets
and of stellar pulsations.
The absence of a pulsation signal in $\tau$ Boo supports the
idea that companions of roughly jovian mass may exist in close
orbits around Sun-like stars.
If one such object circles $\tau$ Boo then it is plausible that
another circles 51 Peg,
questions about the companions' \ process of origin notwithstanding.
As an issue in stellar seismology, the absence of large-amplitude
(by solar standards) pulsations in Sun-like stars would be disappointing
but not surprising.
Our evidence regarding 51 Peg itself is not conclusive,
while the observations by GH raise interesting questions
but lack confirmation.
Further data on this star may therefore be desirable.

We are grateful to the rest of the AFOE team
(Martin Krockenberger and Adam Contos) and to
the staff at SAO's Whipple Observatory (Bastian van't Saant,
Ted Groner, Perry Berlind, Jim Peters, and Wayne Peters)
for their assistance in obtaining and reducing
the observations described here.
We thank Coen Schrijvers for providing test cases against which to
compare our line profile simulation code,
and Artie Hatzes for many useful discussions.

\vskip12pt
\parindent=0pt
{\bf References}
\def\ref{\leftskip20pt \parindent-20pt \parskip4pt}

\ref
Abramowitz, M. \& Stegun, I. 1972, {\it Handbook of Mathematical Functions},
Dover, New York, p. 775

\ref
Brown, T.M., Noyes, R.W., Nisenson, P., Korzennik, S., \& Horner, S. 1994, 
PASP 106, 1285

\ref
Brown, T.M., Kotak, R., Horner, S.D., Kennelly, E.J., Korzennik, S.,
Nisenson, P. \& Noyes, R.W. 1998,
ApJ Supp (submitted)

\ref
Butler, R.P., Marcy, G.W., Williams, E., Hauser, H., \& Shirts, P. 1997,
ApJ Lett. 474, L115

\ref
Gray, D.F. 1997, Nature, 385, 795

\ref
Gray, D.F. \& Hatzes, A.P. 1997,
ApJ (in press)

\ref
Hatzes, A.P. 1996, PASP 108, 839

\ref
Kennelly, E.J., Brown, T.M., Kotak, R., Sigut, T.A.A., Horner, S.D.,
Korzennik, S.G., Nisenson, P., Noyes, R.W., Walker, A., \& Yang, S. 1997,
ApJ (in press)

\ref
Marcy, G.W., Butler, R.P., Williams, E., Bildsten, L., Graham, J.R.,
Ghez, A.M., \& Jernigan, J.G. 1997, ApJ 481, 926

\ref 
Mayor, M. \& Queloz, D. 1995, Nature, 378, 355

\ref
Noyes, R.W., Jha, S., Korzennik, S., Krockenberger, M., Nisenson, P.,
Brown, T.M., Kennelly, E.J., \& Horner, S.D. 1997, ApJ Lett. 483, 111

\ref
Schrijvers, C., Telting, J.H., Aerts, C., Ruymaekers, E., \& Henrichs, H.F.
1997, A\&A Supp. 121, 343

\ref
Vogt, SS. \& Penrod, G.D. 1983, ApJ, 275, 661

\vfill\eject
\baselineskip=20pt

\centerline{\bf Table 1}
\centerline{Observed Amplitudes and Errors for 51 Peg \& $\tau$ Boo}
$$\vbox
{\halign{\hfil#\hfil \qquad & #\hfil \qquad & #\hfil \qquad &
#\hfil \cr
\multispan4\hrulefill \cr
  &51 Peg & 51 Peg & $\tau$ Boo \cr
  & 4.231 d & 2.575 d & 3.313 d \cr
\multispan4\hrulefill \cr
$h_3$ (\%) & 0.147 $\pm$ 0.26 & 0.344 $\pm$ 0.38 & 0.270 $\pm$ 0.35 \cr
$h_4$ (\%) & 0.272 $\pm$ 0.42 & 0.238 $\pm$ 0.45 & 0.148 $\pm$ 0.30 \cr 
$h_5$ (\%) & 0.219 $\pm$ 0.19 & 0.323 $\pm$ 0.19 & 0.190 $\pm$ 0.17 \cr
Line Ratio & 0.277 $\pm$ 0.30 & 0.416 $\pm$ 0.28 &\quad  -- \cr
\multispan4\hrulefill \cr
}}$$

\vfill\eject

{\bf Figure Captions}

\ref
Figure 1.  The line shape parameter $h_3$ for 51 Peg plotted against phase
of the radial velocity variation.
Filled circles are data points covering one cycle of the RV variation;
open circles are the same data covering a wider phase range,
plotted for clarity of the phase relations.
The solid curve is a least-squares sine wave fit to the observations, with
amplitude 0.147\%.
The other curves are the typical $h_3$ signals 
expected from pulsation models
yielding (dashed) an RV signal $V_{dop} = 56$ m s$^{-1}$,
and
(dot-dashed) a line bisector curvature signal of 45 m s$^{-1}$.

\ref
Figure 2.  Same as Fig. 1, but for $\tau$ Boo.
In this case no bisector curvature information is available, so
the corresponding (dot-dashed) curve is not shown.

\vfil\eject

\input psfig
\psfig{figure=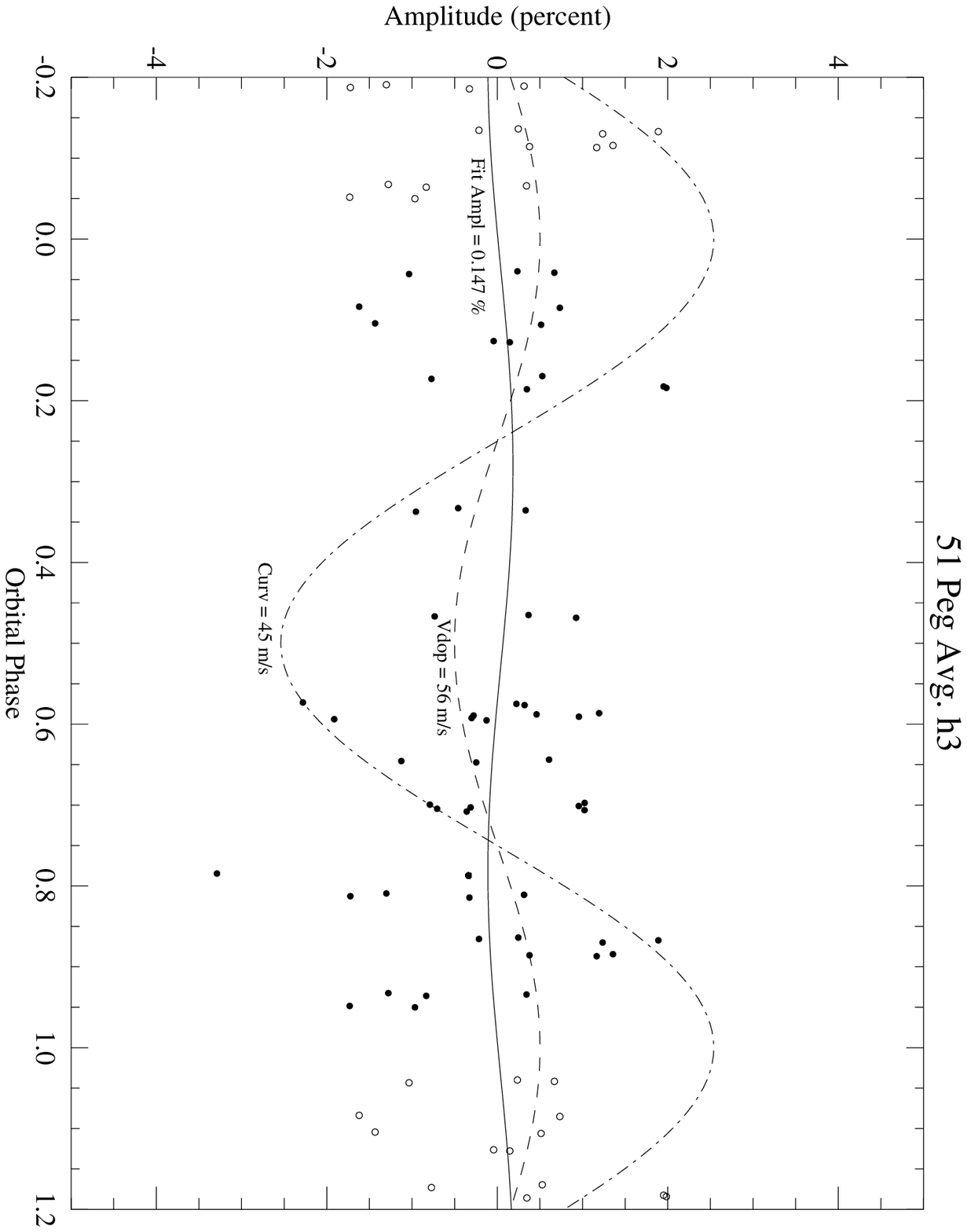}
\psfig{figure=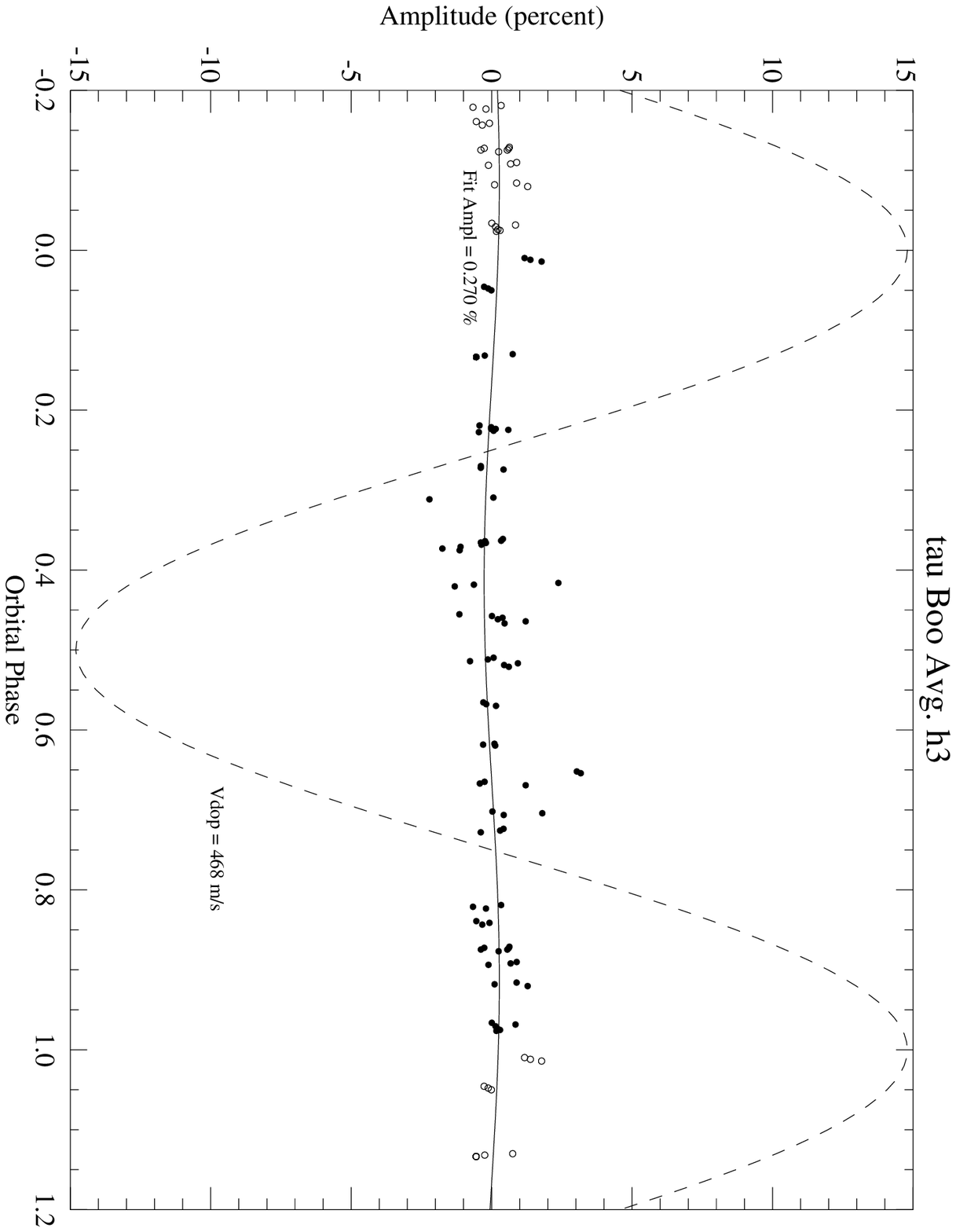}

\bye